\documentclass[useAMS,usenatbib]{mn2e}
\usepackage{graphicx}
\usepackage{txfonts}
\usepackage{natbib}
\title[Blazhko RRc Star LS Her]{The Unique Frequency Spectrum of the Blazhko RRc Star LS Her}
\author[P. Wils, S. Kleidis \& E. Broens]{Patrick Wils$^{1}$, Stelios Kleidis$^{2,3}$, Eric Broens$^{1}$\\
$^{1}$Vereniging voor Sterrenkunde, Belgium, email: patrickwils@yahoo.com,  eric.broens@skynet.be \\
$^{2}$Zagori Observatory, Epirus, Greece, e-mail: steliosklidis@gmail.com \\
$^{3}$Helliniki Astronomiki Enosi, Athens, Greece}

\begin{document}

\date{Accepted 2008 March 29.  Received 2008 March 24; in original form 2008 February 15}

\pagerange{\pageref{firstpage}--\pageref{lastpage}} \pubyear{2008}

\maketitle

\label{firstpage}

\begin{abstract}
The Blazhko effect in RR~Lyrae stars is still poorly understood theoretically.  
Stars with multiple Blazhko periods or in which the Blazhko effect itself varies are particularly challenging.
This study investigates the Blazhko effect in the RRc star LS~Her.
Detailed CCD photometry in the $V$, $R_C$ and $I_C$ band has been performed on 63 nights during six months.
LS~Her is confirmed to have a Blazhko period of $12.75\pm0.02$ days.  
However, where normally the side frequencies of the Blazhko triplet are expected, 
an equidistant group of three frequencies is found on both sides of the main pulsation frequency.
As a consequence the period and amplitude of the Blazhko effect itself vary in a cycle of $109\pm4$ days.
LS~Her is a unique object turning out to be very important in the verification of the theories for the Blazhko effect. 
\end{abstract}


%
%________________________________________________________________

\section{Introduction}

A century after the discovery of the Blazhko effect in RR~Lyrae stars \citep{blazhko}, it can still not be explained theoretically in a satisfactory way \citep{rrlyr}.
The Blazhko effect manifests itself in the frequency spectrum of the light variations 
as one or more side frequencies close to the main pulsation frequency.
The beat period resulting from the interaction between the main frequency and the side frequency is called the Blazhko period.
The question focuses on how many side frequencies there are.
Resonance models involving non-radial pulsations with degree $\ell=1$ lead to an equidistant triplet structure that is most commonly observed: 
two additional frequencies, one on each side of the main frequency and at the same distance.
In some cases only doublets are observed, presumably because the amplitude of the third frequency is too low to be detected.
Oblique magnetic models with $\ell=2$ lead to an equidistant quintuplet structure.
Recently a frequency quintuplet was found for the first time in RV~UMa \citep{rvuma}.
Although admittedly very rare, some RR~Lyrae stars show two Blazhko periods.  
These stars pose a particular challenge for the theories. 

LS~Her is one of the stars known to have two Blazhko periods.  
It was discovered to be variable by \citet{hoffmeister} and first classified as a W~UMa type
eclipsing binary.  Observations on six consecutive nights by \citet{binnendijk} revealed its nature
as an RR~Lyrae star pulsating in the first overtone mode (an RRc type object).  
With a period of 0.2316 days \citep{binnendijk} it has one of the shortest periods among the RRc stars.  
Binnendijk also noted variations in the light curve from night to night, 
but did not find a beat period because of the short timespan of his observations.
From the public survey data of the Northern Sky Variability Survey \citep[NSVS; ][]{nsvs}, 
\citet{rrnsvs} found LS~Her to be a Blazhko star with a Blazhko period of 13 days.
\citet{rrasas} found two pairs of Blazhko triplets resulting in periods of 11.5
and 12.8 days from data of the All Sky Automated Survey \citep[ASAS-3; ][]{asas}.  

Other Galactic examples with multiple Blazhko periods are XZ~Cyg with periods of 57.5 and 41.6 days \citep{xzcyg}, 
UZ~UMa with periods of 26.7 and 143 days \citep{uzuma} 
and SU~Col with periods of 65.8 and 89.3 days \citep{rrasas}.
Unlike LS~Her, all these stars are fundamental mode pulsators (RRab stars).
Until recently, only a few Galactic field Blazhko RRc stars were known \citep{smith, kolenberg}.
Because of its short period it is possible to observe a full pulsation cycle of LS~Her during a night.  
Combined with its double short Blazhko cycles, this made LS~Her a very interesting object for detailed study.

The spectral type of LS~Her has been given as A2 \citep{goetz}, A5 \citep{gcvs}
and F0 \citep{pmcco}.

%__________________________________________________________________

\section{Observations}

LS~Her was observed on 63 nights during six months in 2007 from two private observatories.  
On three nights the star was observed at both observatories.
More than 240 hours of CCD photometry were secured in $V$ and $I_C$, and 65 hours on 22 nights in $R_C$.
The complete observation log, together with information on the instruments used, is given in Table~\ref{log}.  

\begin{table*}
\begin{center}
\caption{Observation log.}
\label{log}
\begin{tabular}{cllcccrccc}
\hline
Observer & Telescope & CCD camera & Timespan & Nbr of & Nbr of & \multicolumn{3}{c}{Number of data points} \\
initials &           &            &  {\small (JD-2450000)} &  nights & hours & $V$ & $R_C$ & $I_C$ \\
\hline
SK    & 30-cm LX200 & SBIG ST-7XMEI & 4199-4362 & 44 & 181.1 & 3613 & -   & 3619 \\
EB    & 20-cm C8    & SBIG ST-7XMEI & 4206-4366 & 22 &  65.5 &  899 & 901 &  892 \\
Total &             &               & 4199-4366 & 63 & 243.2 & 4512 & 901 & 4511 \\
\hline
\end{tabular}
\end{center}
\end{table*}

\begin{figure*}
\centering
\includegraphics[width=17cm]{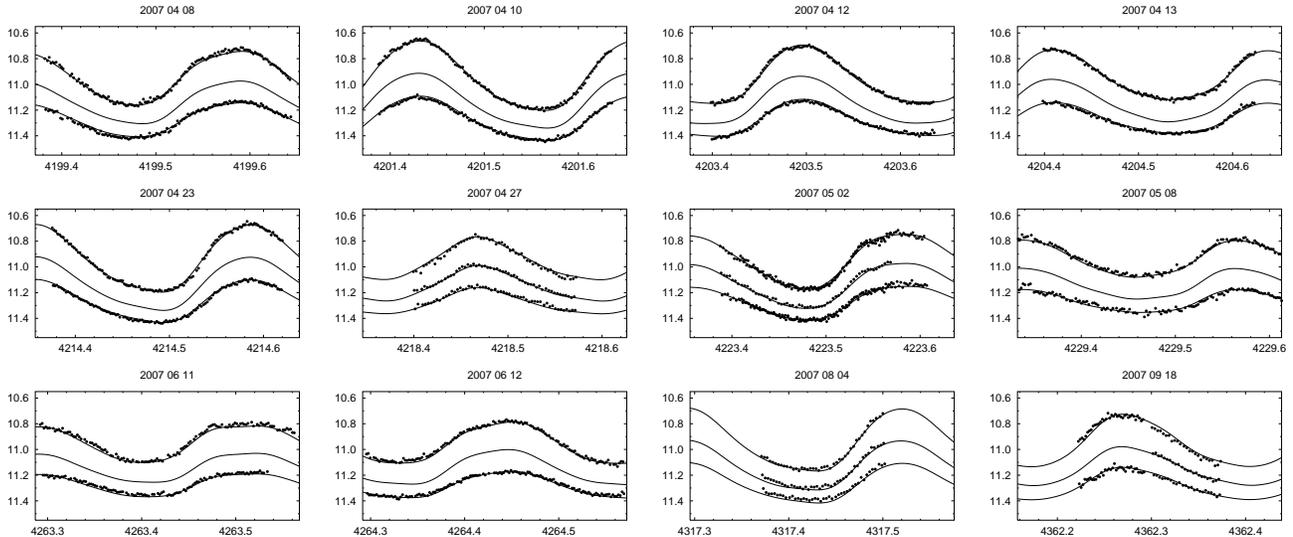}
\caption{$V$ (top curve), $R_C$ (middle) and $I_C$ (bottom curve) data of LS~Her on a number of nights.  
The zero points of the $R_C$ and $I_C$ magnitude scales have been arbitrarily chosen.
Labels on the horizontal axis are $HJD-2450000$.
The model plots from Table~\ref{freq} are shown as well.}
\label{plot}
\end{figure*}

Differential photometry was performed using GSC~1507-598 or GSC~1507-660 as comparison star 
and GSC~1507-776 or GSC~1507-327 as check star.
Nightly standard deviations on the difference between the check star and comparison star were usually around 0.01 magnitude or better.
All magnitudes were reduced to a common scale assuming the difference in magnitude between GSC~1507-660 and GSC~1507-598 
to be $\Delta V = -0.446$, $\Delta R_C = -0.254$ and $\Delta I_C = -0.060$ (determined from our data on 13 nights).  
GSC~1507-598 was then further assumed to have $V = 12.16$ from ASAS-3.
No instrumental corrections were applied to the data.
Some nightly plots are shown in Fig.~\ref{plot}, together with the model fit derived in the next section.

%__________________________________________________________________

\section{Frequency analysis}

Period04 \citep{period04} was used for the frequency analysis.  
Table~\ref{freqsurveys} gives the frequencies found in the NSVS and ASAS-3 survey data.  
It lists the value of the frequency, its semi-amplitude in mmag and the signal to noise ratio ($S/N$).
Uncertainties on the given values are the standard deviations from the results obtained from Monte Carlo simulations.

\begin{table*}
\begin{center}
\caption{Frequencies for LS~Her found in NSVS and ASAS-3 data.}
\label{freqsurveys}
\begin{tabular}{clrrlrr}
\hline
 & \multicolumn{3}{c}{NSVS} & \multicolumn{3}{c}{ASAS-3} \\
Ident. & Frequency & \multicolumn{1}{c}{$A_{NSVS}$} & $$S/N$$ & Frequency & \multicolumn{1}{c}{$A_V$} & $$S/N$$ \\
 & \multicolumn{1}{c}{(c/d)}  & \multicolumn{1}{c}{(mmag)} & & \multicolumn{1}{c}{(c/d)}  & \multicolumn{1}{c}{(mmag)} \\
\hline
$f_0$       & 4.33262(4) & 150(4) & 31.5 & 4.33258(1)  & 181(5) & 33.5 \\
$f_0 + f_1$ & 4.41097(20) & 45(4) &  9.3 & 4.41088(8)  &  58(5) & 10.8 \\
$f_0 - f_1$ & 4.25426(20) & 27(4) &  5.6 & 4.25429(8)  &  45(5) &  8.4 \\
$f_0 + f_2$ & 4.42009(20) & 20(4) &  4.3 & 4.42090(30) &  22(7) &  4.2 \\
$f_0 - f_2$ & 4.24514(20) & 33(4) &  6.8 & \multicolumn{1}{c}{-} & \multicolumn{1}{c}{-} & \\
\hline
\end{tabular}
\end{center}
\end{table*}

\begin{figure}
\centering
\includegraphics[width=8.5cm]{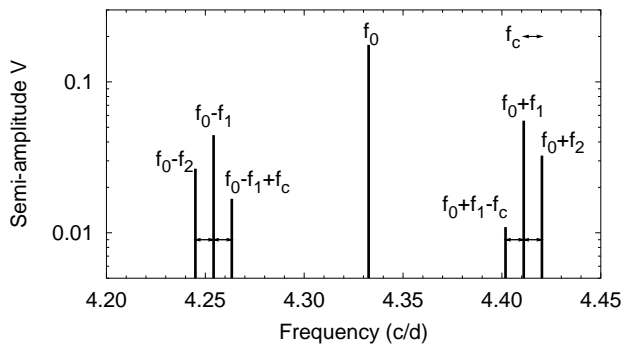}
\caption{Schematic diagram of the observed frequencies around the main pulsation frequency.}
\label{freqs}
\end{figure}

The five frequencies listed in Table~\ref{freqsurveys} constitute in fact two triplets around the main frequency which is common to both triplets.
The two triplets were also found in our own data sets.
The Blazhko (or beat) period resulting from the first triplet is $12.75\pm0.02$ days and $11.42\pm0.04$ days from the second triplet.
Surprisingly however our data also revealed a third triplet, again with the side frequencies close to those of the first triplet 
but on the other side of those of the second triplet and at about the same distance.  
This third triplet corresponds to a Blazhko period of $14.45\pm0.07$ days.
A schematic presentation of these frequencies is given in Fig.~\ref{freqs}.  
A logarithmic scale is used for the $V$ amplitudes.

Due to the proximity of all the side frequencies and the limitations of the procedure to find the next significant frequency in the data,
it was impossible to find a good frequency fit while imposing equidistance conditions from the start 
(see \citet{equidistance} for a discussion on the equidistance of the triplets).  
Instead, a satisfactory solution could only be found by leaving all frequencies independent of each other.  
In the end, when seven independent frequencies were found, it was clear that  
the distances to the main pulsation frequency of the frequencies on both sides of the main frequency in each triplet 
were equal within the uncertainties.
For example the distance to the main frequency for the most significant side peak to the right of the main frequency was found to be 
$f_{1+}=0.07840\pm0.00009$ cycles per day (c/d), while for the peak on the left it was found to be $f_{1-}=0.07847\pm0.00006$ c/d.
Similarly the distance between the side frequencies of the first and second triplet was found to be $f_{c+}=0.0092\pm0.0007$ c/d,
and the distance for those of the first and third triplet was $f_{c-}=0.0093\pm0.0010$ c/d.
The final frequency solution was then obtained using the equidistance criterion and is listed in Table~\ref{freq}.
Only frequencies with $S/N > 4$ were retained \citep{breger}. 
In addition to amplitudes and signal to noise ratio in $V$, $R_C$ and $I_C$, 
the phase difference of the frequencies between $V$ and $I_C$ are given.  
The latter may aid in mode identification \citep[see e.g. ][ and references therein]{dasz}.
The small positive phase difference for the radial mode $f_0$ is consistent with that found for other RR~Lyrae variables \citep[e.g. RV~UMa, ][]{rvuma}.
Only one of the six side frequencies has a phase difference different from zero at the 1-$\sigma$ level, 
but it is not significant at the 2-$\sigma$ level.

\begin{table*}
\begin{center}
\caption{Frequencies for LS~Her detected in our data sets. }
\label{freq}
\begin{tabular}{cr@{}lr@{}lrr@{}lrr@{}lrr}
\hline
\multicolumn{3}{c}{Frequency} & \multicolumn{2}{c}{$A_V$} & \multicolumn{1}{c}{$S/N$} 
                              & \multicolumn{2}{c}{$A_R$} & \multicolumn{1}{c}{$S/N$} 
                              & \multicolumn{2}{c}{$A_I$} & \multicolumn{1}{c}{$S/N$} 
                              & \multicolumn{1}{r}{$\Phi_V - \Phi_I$} \\
Identification & \multicolumn{2}{c}{(c/d)}  
                              & \multicolumn{2}{c}{(mmag)} & \multicolumn{1}{c}{$V$}
                              & \multicolumn{2}{c}{(mmag)} & \multicolumn{1}{c}{$R_C$}
                              & \multicolumn{2}{c}{(mmag)} & \multicolumn{1}{c}{$I_C$}
                              & \multicolumn{1}{r}{(degrees)} \\
\hline
$f_0$ & 4.33261 & (2) & 176.3 & (4) & 147.0 & 144.7 & (10) & 73.2 & 108.2 & (4) & 90.1 & $2.6\pm0.3$ \\
$f_0+f_1$ & 4.41101 & (15) & 55.4 & (7) & 46.6 & 48.0 & (16) & 24.6 & 35.3 & (4) & 29.3 & $-0.6\pm1.1$ \\
$f_0-f_1$ & 4.25421 &  & 44.4 & (5) & 36.4 & 26.3 & (18) & 13.1 & 28.3 & (4) & 23.2 & $0.9\pm1.1$ \\
$f_0+f_1+f_c$ & 4.42021 & (34) & 32.5 & (5) & 27.4 & 22.5 & (17) & 11.5 & 20.5 & (4) & 17.0 & $0.4\pm1.6$ \\
$f_0-f_1-f_c$ & 4.24501 &  & 26.6 & (5) & 21.8 & 23.0 & (17) & 11.5 & 16.6 & (4) & 13.6 & $1.0\pm2.1$ \\
$f_0+f_1-f_c$ & 4.40181 &  & 10.9 & (6) & 9.1 &  & &  & 7.5 & (4) & 6.2 & $2.1\pm5.0$ \\
$f_0-f_1+f_c$ & 4.26341 &  & 16.8 & (10) & 13.8 & 17.2 & (14) & 8.6 & 10.4 & (4) & 8.6 & $7.5\pm3.8$ \\
$2f_0$ & 8.66522 &  & 16.5 & (4) & 18.0 & 15.9 & (11) & 11.9 & 11.8 & (5) & 14.1 & $4.6\pm2.5$ \\
$2f_0+f_1$ & 8.74362 &  & 13.5 & (4) & 14.6 & 11.4 & (11) & 8.6 & 9.6 & (4) & 11.5 & $-11.4\pm3.3$ \\
$2f_0-f_1$ & 8.58681 &  & 10.0 & (4) & 10.8 & 8.9 & (11) & 6.7 & 6.8 & (4) & 8.1 & $4.4\pm3.7$ \\
$2f_0+f_1+f_c$ & 8.75282 &  & 8.4 & (5) & 9.1 &  & &  & 4.7 & (4) & 5.6 & $5.4\pm5.5$ \\
$2f_0-f_1-f_c$ & 8.57762 &  & 6.0 & (4) & 6.5 &  & &  &  & &  &   \\
$2f_0+2f_1$ & 8.82202 &  & 5.0 & (4) & 5.4 &  & &  & 3.3 & (4) & 4.0 & $-7.1\pm8.4$ \\
$3f_0$ & 12.99783 &  & 6.9 & (6) & 7.7 & 6.0 & (9) & 4.8 & 4.5 & (4) & 6.3 & $2.8\pm6.4$ \\
$3f_0-f_1$ & 12.91942 &  & 4.7 & (4) & 5.3 &  & &  &  & &  &   \\
$4f_0$ & 17.33043 &  & 4.6 & (5) & 6.1 &  & &  & 2.9 & (4) & 4.8 & $0.3\pm9.5$ \\
\hline
\end{tabular}
\end{center}
\end{table*}

\begin{figure}
\centering
\includegraphics[width=8cm]{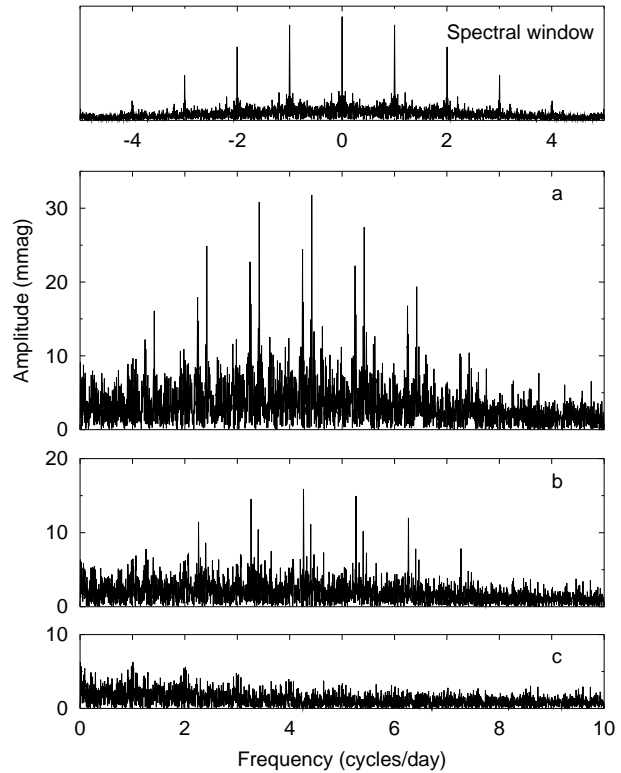}
\caption{Frequency spectrum of our $V$ data for LS~Her.  The top panel shows the spectral window.  
Panels a, b and c show the spectrum after consecutively prewhitening for the first, second and third triplet respectively.}
\label{fourier}
\end{figure}

\begin{figure}
\centering
\includegraphics[width=7.5cm]{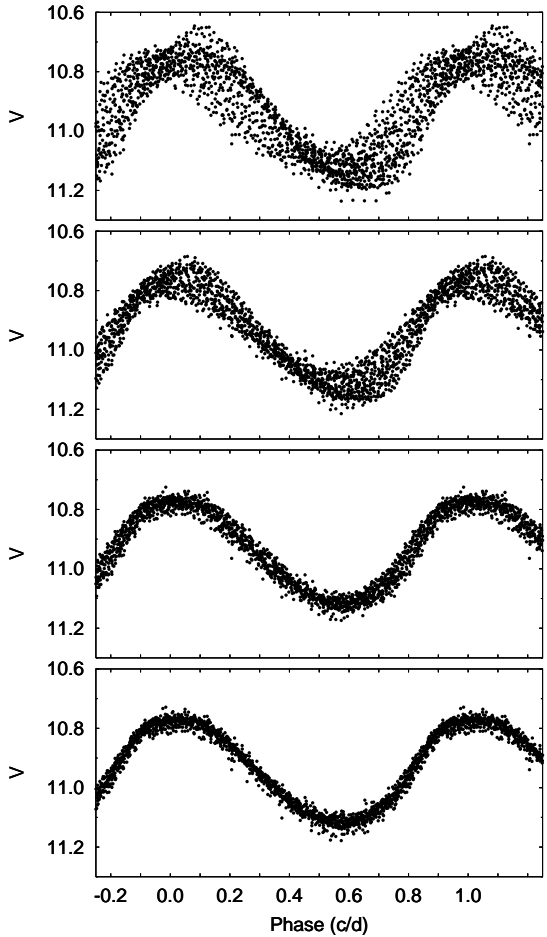}
\caption{Phase diagrams of our $V$ data for LS~Her plotted with the main pulsation period of 0.230808 days.
From top to bottom the data are plotted without prewhitening and then with consecutively prewhitening for the first, second and third triplet.
In order not to overload the diagrams, only one out of three data points have been plotted.}
\label{phase}
\end{figure}

Fig.~\ref{fourier} shows several frequency spectra created from the $V$ data.  
Besides the spectral window it shows the frequency spectrum after removal of the first, second and third triplet side frequencies.
It can be seen that after removing all frequencies given in Table~\ref{freq}, 
there seems to be some power left at frequencies near integer fractions of a day.
The amplitude is however small, less than 6 mmag.  
This remaining power is possibly caused by small zero point shifts in the data from night to night,
likely introduced by the corrections needed to account for the use of different comparison stars.
Phase diagrams after consecutive cleaning of our $V$ data with the three frequency triplets are given in Fig.~\ref{phase}.
  
The second and third triplets may be viewed as changing the (primary) Blazhko modulation itself. 
With a period of $109\pm4$ days the amount of variation in amplitude and phase will change.
In the frequency spectrum this manifests itself by the side frequencies of the first Blazhko triplet being equidistant triplets themselves.
In Table~\ref{freq} we have therefore dubbed $f_c = f_2 - f_1$ as the frequency of this superposed modulation.

\begin{figure}
\centering
\includegraphics[width=8cm]{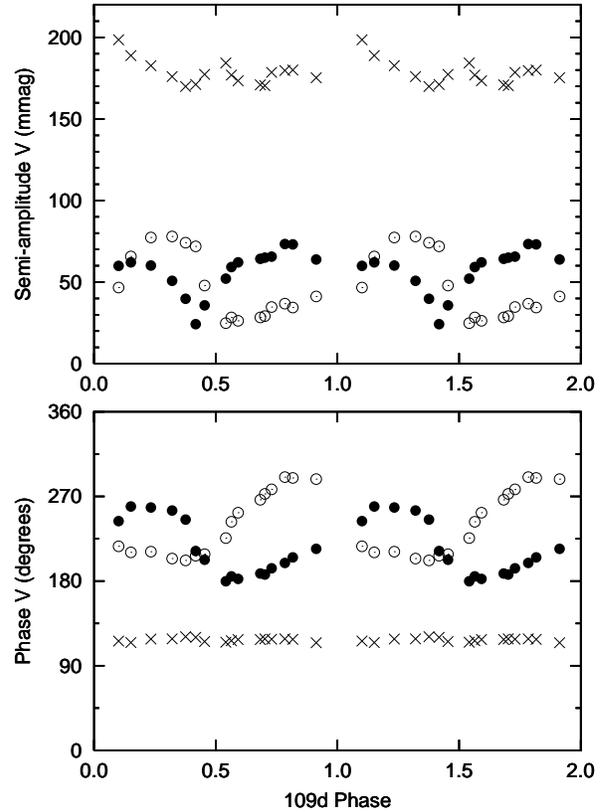}
\caption{Changes of the amplitude (top panel) and phase (bottom panel) during the long cycle of 109 days of the main pulsation frequency $f_0$ (crosses) and the Blazhko frequencies
$f_0+f_1$ (filled circles) and $f_0-f_1$ (open circles).}
\label{change}
\end{figure}

To illustrate the modulation of the Blazhko period, an analysis was done of small subsets of the $V$ data set.
Each subset was taken to be 25 days in length (about two Blazhko cycles), 
and the best fitting amplitudes and phases were determined for the pulsation frequency $f_0$, 
the main Blazhko frequencies $f_0+f_1$ and $f_0-f_1$, and their linear combinations.
Precise determination of the frequencies themselves is rather sensitive to the distribution of the data in such a small set.
Therefore the frequencies were kept fixed at the values determined from the total dataset.  
A change in their value would then be manifested as a change in their phase.
The results of this analysis are shown in Fig.~\ref{change}.  
The top panel shows the change in amplitude of $f_0$, $f_0+f_1$ and $f_0-f_1$, as a function of the phase in the 109-day cycle.
The lower panel of Fig.~\ref{change} shows the change in phase.  
As there is no change in phase of the main pulsation period, it remains effectively constant during the long cycle.
However, the Blazhko period appears to vary cyclically in that interval, as can be seen from the phase changes of the Blazhko frequencies.

The variation in the Blazhko modulation may perhaps be compared to the four year cycle in the Blazhko period of RR~Lyr \citep{detre}.  
However, possible splitting of the side frequencies similar to what is observed in LS~Her, has not yet been detected in RR~Lyr, perhaps because of this long period and the very long data sets needed.  
A phase shift of the Blazhko maximum as seen at the end of a 4-year cycle in RR~Lyr was not observed in LS~Her.

As can be seen in Table~\ref{freq}, the second order triplet frequencies (around the first harmonic of the main pulsation frequency) 
are found for the first ($f_1$) and second ($f_1+f_c$) triplet, but not for the third ($f_1-f_c$), presumably because their amplitude is too low.
There is no indication of the quintuplet frequencies $f_0\pm2f_1$,   
However, the frequency $2f_0+2f_1$ has a significant amplitude in the data.

%
%______________________________________________________________

\section{Conclusion}

LS~Her was found to be an RRc Blazhko star with three close Blazhko triplets around the main pulsation frequency.
The side frequencies of each of the three triplets are at the same distance from the main first overtone pulsation frequency.
In addition, the side frequencies of the second and third triplets are at the same distance from the side frequencies of the first triplet, within uncertainties.
Consequently the maximum amplitude and maximum phase shift of the Blazhko effect changes.
The Blazhko effect was found to have a primary period of 12.75 days, while its influence on amplitude and phase shifts, due to the additional triplets, changes in a cycle of 109 days.

The frequency spectrum of LS~Her has been found to be unique so far, but it is not excluded that other stars 
which are known to have two Blazhko periods also show this complex frequency structure.
As their Blazhko periods are much longer, the missing triplet would also be much harder to detect.
Adequate coverage of a number of consecutive Blazhko cycles would be required to reveal additional frequencies in the spectrum of a Blazhko star.

The complex frequency spectrum of LS~Her cannot easily be inferred from the current theories of the Blazhko effect.
LS~Her is therefore an important star against which to verify any theory of the Blazhko effect.

%
%______________________________________________________________

\section*{Acknowledgements}
The authors gratefully acknowledge Katrien Kolenberg for helpful suggestions, 
Brian Skiff for providing the spectral type references, and the referee for his valuable comments.
EB acknowledges the AAVSO and the Curry Foundation for providing the CCD camera and filters on loan.
This study used data from the Northern Sky Variability Survey 
created jointly by the Los Alamos National Laboratory and the University of Michigan,
and funded by the US Department of Energy, 
the National Aeronautics and Space Administration (NASA) and the National Science Foundation (NSF).
This study also used NASA's Astrophysics Data System, and the SIMBAD and VizieR 
databases operated at the Centre de Donn\'ees Astronomiques (Strasbourg) in France.

\label{lastpage}

\end{document}